\documentclass[lettersize,journal]{IEEEtran}
\usepackage{mathptmx}
\usepackage{times}
\IEEEoverridecommandlockouts
\usepackage{graphicx,color,float}
\usepackage{amssymb,amsmath,times,wasysym,comment}
\usepackage{cite}
\usepackage{epsfig}
\UseRawInputEncoding
\usepackage{diagbox}
\newcolumntype{C}{>{\centering\arraybackslash}X}
\usepackage[hyphens]{url}
\usepackage{hyperref}
\usepackage{acronym}
\usepackage{nicematrix}
\usepackage{xcolor}
\hypersetup{breaklinks=true}

\def\BibTeX{{\rm B\kern-.05em{\sc i\kern-.025em b}\kern-.08em
		T\kern-.1667em\lower.7ex\hbox{E}\kern-.125emX}}
	\def\BibTeX{{\rm B\kern-.05em{\sc i\kern-.025em b}\kern-.08em
	T\kern-.1667em\lower.7ex\hbox{E}\kern-.125emX}}

\acrodef{QSDC}{quantum superdense coding}
\acrodef{QKD}{quantum key distribution}
\acrodef{3GPP}{3rd Generation Partnership Project}

\title{Digital Twin for Non-Terrestrial Networks: Vision, Challenges, and Enabling Technologies}

 \author{\normalsize Hayder Al-Hraishawi,~\IEEEmembership{Senior Member,~IEEE}, Madyan Alsenwi,~\IEEEmembership{Member,~IEEE},\\ Junaid ur Rehman, Eva Lagunas,~\IEEEmembership{Senior Member,~IEEE}, and Symeon Chatzinotas, ~\IEEEmembership{Fellow,~IEEE}\\

\thanks{The authors are with the Interdisciplinary Centre for Security, Reliability and Trust (SnT), University of Luxembourg, Luxembourg.
Corresponding author: \textit{Hayder Al-Hraishawi (hayder.a@siu.edu)}.}

\vspace{-5mm}} 

\begin{document}

\bstctlcite{IEEEexample:BSTcontrol}
\maketitle
\thispagestyle{plain}
    \pagestyle{plain}	
   
\begin{abstract}

	This paper investigates the transformative potential of digital twin (DT) technology for non-terrestrial networks (NTNs). NTNs, comprising airborne and space-borne elements, face unique challenges in network control, management, and optimization. DT technology provides a novel framework for designing and managing complex cyber-physical systems with enhanced automation, intelligence, and resilience. By offering a dynamic virtual representation of the NTN ecosystem, DTs enable real-time monitoring, simulation, and data-driven decision-making. This paper explores the integration of DTs into NTNs, identifying technical challenges and highlighting some key enabling technologies. Emphasis is placed on technologies such as the Internet of Things (IoT), machine learning, generative AI, space-based clouds, quantum computing, and others, highlighting their potential to empower DT development for NTNs. To illustrate these concepts, we present a case study demonstrating the implementation of a data-driven DT model for enabling dynamic, service-oriented network slicing within an open radio access network (O-RAN) architecture tailored for NTNs. This work aims to advance the understanding and application of DT technology, contributing to the evolution of network control and management in the dynamic and rapidly changing landscape of non-terrestrial communication systems. 
\end{abstract}


\section{Introduction}\label{sec:intro}
In light of the recent  technological advancements in non-geostationary orbit (NGSO) satellites, high altitude platforms (HAPs), and unmanned aerial vehicles (UAVs), the creation of non-terrestrial networks (NTNs) has become a tangible reality \cite{al2022non}. NTNs offer practical solutions for providing communication services to remote and rural areas, addressing the limitations of traditional terrestrial networks in these regions \cite{Hayder2021}. Thus, NTNs have a crucial role in ensuring global coverage for applications requiring high availability and resilience, as well as bridging the digital divide, fostering innovation, stimulating economic growth, and driving social progress. However, these benefits of NTNs come at the expense of higher complexity due to the massive number of network entities and users that are interacting within a dynamic and heterogeneous environment. Hence, the design and management of NTNs become increasingly costly and difficult.

To address these complexities, \emph{digital twin (DT)}  technology emerges as a promising solution, offering a detailed virtual representation of the NTN ecosystem. DT-NTNs continuously integrate real-time data, enabling advanced simulations to test network behavior, proactive monitoring to address technical issues, and data-driven decision-making to optimize performance and resource allocation \cite{Mihai2022, Mohammadi2021}. For instance, DT-NTNs can enhance energy efficiency by leveraging historical traffic data to dynamically adjust power levels of network elements during periods of low activity. This real-time adaptability ensures optimal performance while conserving resources. Furthermore, DTs accelerate NTN innovation by providing a precise emulation environment for testing new technologies and features before deployment, reducing investment risks and fostering reliable advancements in non-terrestrial communication systems \cite{Zhou2023}.

\begin{figure*}[!t]
	\centering
	\def\svgwidth{500pt}
	\fontsize{8}{4}\selectfont{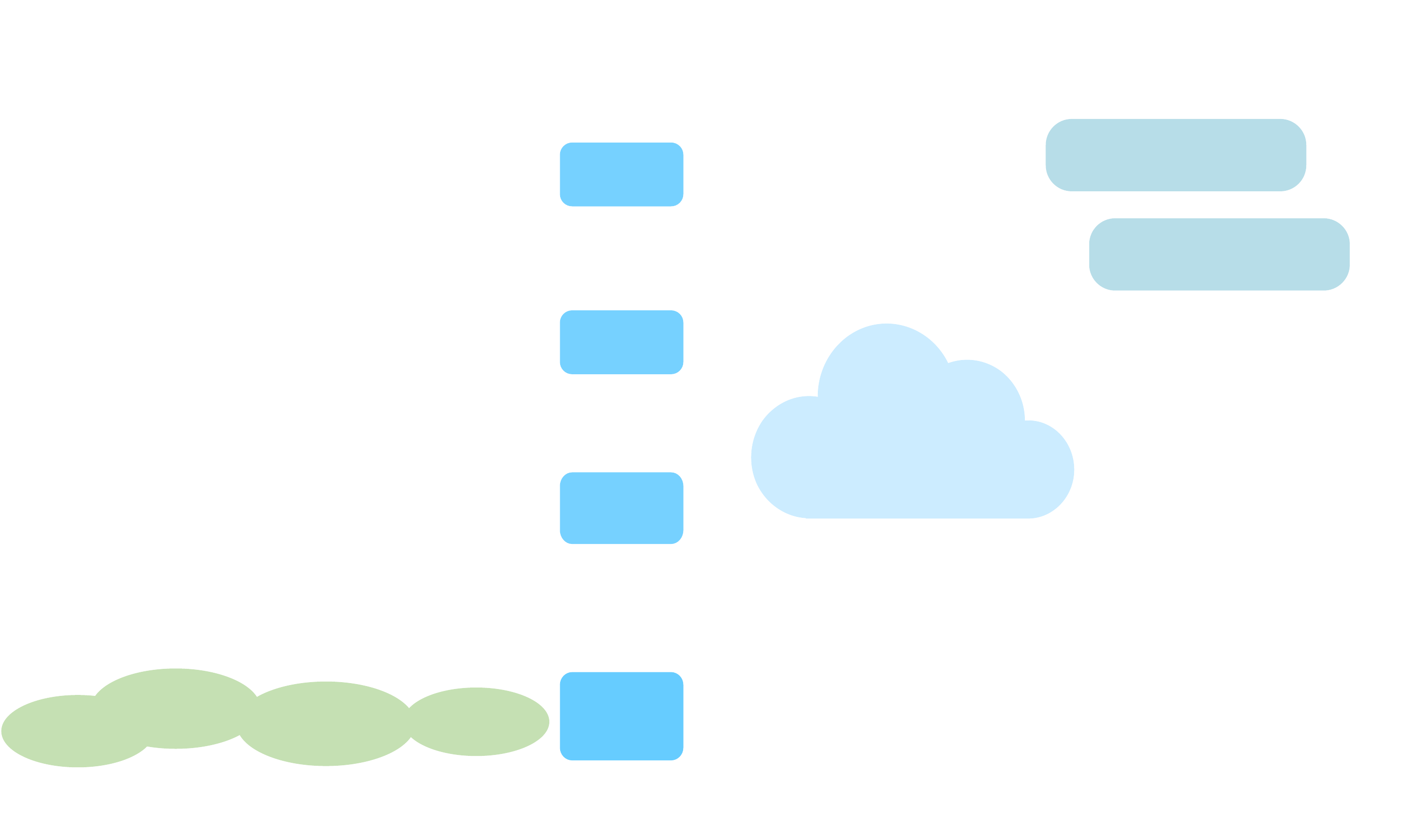}
	\caption{General overview of a DT-NTN model.} \label{fig:NTN_DT} 
\end{figure*}

In contrast to traditional simulation tools, DT-NTNs leverage real-time data generated by sensors attached to physical objects to enable online control and algorithmic decision-making \cite{Khan2022}. Further, this two-way information exchange between the DT-NTNs and sensors contributes to a real-time, comprehensive understanding into the performance and behavior of NTN assets. Beyond its virtual representation, DT-NTN models can utilize tools from optimization theory, game theory, and artificial intelligence (AI) to aid in instantaneous monitoring, optimization, and control. Moreover, DT-NTNs  can also function as robust research and development (R\&D) platforms, supporting the creation of innovative network architectures and standards. This multifaceted role of  DT-NTNs emphasizes their substantial value in advancing NTNs. Therefore, this paper investigates  the essential aspects of realizing DT technology within NTNs, identifying the deployment challenges, and exploring the key enabling technologies that support the implementation of efficient  DT-NTNs.

\section{Overview of NTNs and Integration with DT Technology}
According to the \ac{3GPP} specifications, NTN is defined as an umbrella term for communication networks that involve non-terrestrial flying objects including space-borne vehicles such as geostationary earth orbit (GEO), medium earth orbit (MEO), and low earth orbit (LEO) satellites, or airborne vehicles, i.e. HAPs and UAVs \cite{NGSO_survey}. The communication architecture of an NTN typically consists of three segments:
\begin{enumerate}
	\item Space-aerial segment: Includes satellites, HAPs, and UAVs.
	\item Ground segment: Comprising ground stations/gateways that relay data to and from the space-aerial segment.
	\item User segment: Consists of terminals such as ships, airplanes, other ground-based users, and mobile devices.
\end{enumerate}
 The ground segment provides real-time control and management for the communications between different NTN entities through  the network control center (NCC) and  network management center (NMC). The NCC is responsible for the overall instantaneous management and control of NTNs, while the NMC is in charge of monitoring and managing the performance and health of the network elements. This inherent complexity of NTNs has driven research and industry efforts to explore the integration of DT technology.

The concept of DT has long been applied in the satellite research field. For instance, NASA's 2010 technological roadmap highlights some of the ways in which DT is being utilized in aerospace including satellites. Specifically, DT can be used in simulating an actual satellite before launching to maximize the mission success. Besides, DT can be used to continuously mirror an flying object and update its conditions as well as diagnosing any damage. This approach results in extended life expectancy of the satellites and allows for in-situ repairs or mission modifications when necessary \cite{kim2022survey}. 
Beyond aerospace, wireless communication standardization bodies, such as the International Telecommunication Union (ITU) \cite{ITU3090}, are working on developing DT technology and defining its core concepts and interfaces.

\subsection{DT-NTN System Model }
Figure \ref{fig:NTN_DT} presents a high-level architecture of an integrated DT-NTN system model  with three layers

\begin{enumerate}
	\item {Physical Layer}: Encompasses all physical components of the NTN.
	
	\item {Twin Layer}: Hosts virtual representations of physical objects or network states, utilizing data-driven models to simulate traffic, topology, routing algorithms, scheduling schemes, etc.
	\item {Application Layer}: Provides interfaces for operators to monitor, analyze, and optimize performance, offering versatile applications across industries.
\end{enumerate}



An essential element in the development of DT-NTNs is the establishment of efficient and reliable interfaces between their diverse layers and entities. These interfaces can take several forms, including twin-to-physical object, twin-to-twin, and twin-to-service. Two key interfaces are critical to the functioning of NTNs. The first interface collects TT\&C data from flying assets, which is essential for system health and control. The second interface collects communication-related data, such as traffic demands, channel states, topological routes, and connection/failure incidents, allowing effective network management to meet user demands. 

\begin{figure*}[!t]
	\centering
	\def\svgwidth{420pt}
	\fontsize{8}{4}\selectfont{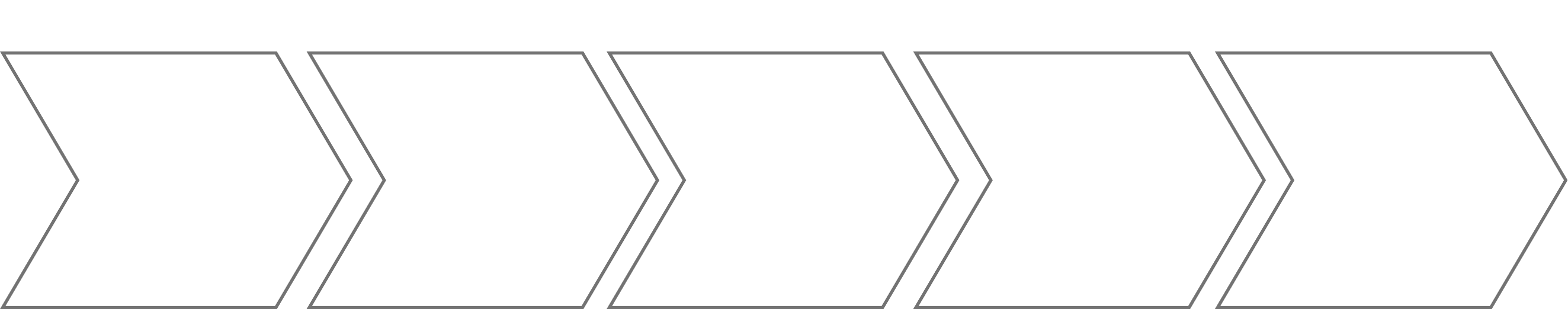}
	\caption{Flowchart depicting the key steps for building DT-NTN models.} \label{fig:Flowchart} 
\end{figure*}
\subsection{DT-NTN Development Steps}
Building a  DT-NTN involves several key steps, each crucial for creating an accurate and functional digital representation of the physical network and its operations. The process can be outlined as follows:
\begin{enumerate}
	\item {Data Gathering}: The first step involves continuous data collection from  NTN assets, including satellites, HAPs, UAVs, ground stations, and user terminals. This data includes telemetry, tracking, and control (TT\&C) data, communication network data, and IoT sensor data.
	
	\item {Data Preprocessing and Integration}
	Collected data from various NTN sources may have inconsistencies, noise, or compatibility issues. Before feeding it into the DT-NTN, preprocessing ensures data quality and consistency. Specifically, data cleansing, normalization, and integration are applied to unify data from different assets. This step helps in reducing errors and improving the accuracy of DT-NTN training.
	
	\item {Model Calibration and Validation}:  The DT-NTN continuously refines digital models of network assets using the gathered data. Calibration and validation steps ensure that these models accurately reflect the real-world behavior, performance, and interactions of each entity within the NTN. By periodically adjusting the models with real-time data, historical performance, and feedback, the DT maintains high accuracy. Validation tests further align the DT with the physical network, enhancing both predictive and adaptive capabilities.

	\item {Processing and Simulation}: Advanced simulation techniques replicate the behavior of the physical NTN in the DT environment. These simulations can predict network performance under various conditions, identify potential issues, and test different resource allocation strategies. 
	
	\item {Post-Operation Analytics}: A continuous feedback loop is established between the DT-NTN and the physical network, ensuring dynamic synchronization with real-time NTN conditions. Through this loop, the DT-NTN provides actionable insights and recommendations to optimize network operations and address emerging challenges. After each operational cycle, the outcomes are analyzed to extract valuable insights, which are then used to refine algorithms and update digital models. This iterative process enhances the DT-NTN's predictive capabilities, making it increasingly accurate and adaptive over time.
\end{enumerate}
This DT-NTN framework ensures efficient management and optimization of the complex and dynamic NTN environment, providing significant enhancements in performance, reliability, and service quality.

\section{Open Challenges}
While DT technology offers clear benefits and enhancements, its implementation within NTNs faces several technical challenges.
 
\begin{itemize}
    \item \textbf{Data Freshness}: This is particularly challenging due to the heterogeneity and dynamicity of NTNs as they operate in remote and harsh environments, where data collection and real-time transmissions may experience delays. While it is feasible to collect and use data for offline operations, achieving online optimization through DTs needs further enhancing real-time data processing capabilities.

    \item \textbf{Ownership and Privacy Concerns}: This challenge arises from the differences in ownership of the diverse entities within NTNs, exacerbated by the General Data Protection Rules (GDPR) in the European Union. Hence, it is imperative to establish well-defined guidelines regarding ownership, data sharing, and data protection in order to ensure seamless operations of DT-NTNs while adhering to GDPR regulations. 

    \item \textbf{Computational Complexity}: The complex and ever-changing characteristics of NTNs require sophisticated modeling techniques. For instance, parameters like flight dynamics, autonomous navigation trajectories, constellation patterns, and communication link performance need to be  modeled as accurately as possible. This requires substantial computational resources and access to high-performance computing infrastructure. Besides, many resource allocation problems in NTNs are inherently non-convex or involve challenging combinatorial optimization, demanding high computational power.

    \item \textbf{Resource-Limited Devices}: NTN entities, such as LEO small satellites, downstream nano-satellites (nanosats), and HAPs, pose a significant challenge due to their constrained computational capacity, limited memory, and often restricted power supply. These devices encounter difficulties in processing and managing the extensive data required for effective digital twining. Therefore, it is imperative to alleviate their burden by offloading storage and computing capabilities. This measure is essential for maintaining sustained functionality and extending their operational lifespan in challenging NTN environments.

   \item \textbf{Interoperability and Standardization}: NTNs often incorporate equipment and systems from various vendors with incompatible configurations, which imposes a challenge for seamless integration and operational efficiency. The lack of interoperability and standardization among vendor equipment further complicates the modeling of NTN element interactions. Therefore, DT-NTNs must be able to collect and integrate data from diverse, autonomous, and heterogeneous sources.
        
    \item \textbf{Security and Reliability}: For an accurate representation of the physical network in the virtual DT-NTNs, secure and reliable communication channels are essential. Any interruption or tampering in these channels can result in inaccurate or incomplete data, compromising the quality of analysis and decision-making. Therefore, robust communication protocols and security measures are crucial to safeguard against such issues and to maintain the DT-NTN's fidelity to the physical network.
\end{itemize}

\section{Enabling Technologies}
In this section, we outline essential enabling technologies along with their salient features pertinent to the development of DT-NTNs. In particular, Table \ref{tab:comparison} presents a summary of key enabling technologies and their potentials, offering effective solutions to surmount DT-NTN deployment challenges. While some of these technologies have been previously discussed  in the framework of NTNs, this discussion offers a fresh perspective on how these advancements can enhance DT functionality within NTNs. By addressing challenges such as computing, sensing, and interoperability, these technologies enable DT-NTNs to dynamically adapt to changing conditions, optimizing resource allocation far more efficiently than traditional methods.

\begin{table*}[]
\caption{Comparison of DT enabling technologies and targeted challenges.} \label{tab:comparison}
\renewcommand{\arraystretch}{1.2}
\begin{tabular}{|l|l|l|l}
\cline{1-3}
\textbf{Enabling Technology}  
&\textbf{Potential}
&\textbf{Targeted Challenge} & \\ \cline{1-3} Narrow-band Internet of things (NB-IoT) & \begin{tabular}[c]{@{}l@{}}Real-time data collection, asset tracking, \\ condition monitoring, predictive maintenance, \\ mission simulation \end{tabular} & Data freshness &       \\ \cline{1-3}
\begin{tabular}[c]{@{}l@{}}Machine learning techniques\end{tabular} & \begin{tabular}[c]{@{}l@{}}Dynamic resource allocation and optimization, \\improved decision-making, anomaly detection \\ failure prediction \end{tabular} & \begin{tabular}[c]{@{}l@{}}Computational complexity\end{tabular}  &       
        \\ \cline{1-3}
\begin{tabular}[c]{@{}l@{}}Generative AI\end{tabular}  & \begin{tabular}[c]{@{}l@{}}Synthetic data generation, data augmentation for training models,\\ scenario simulation and testing, enhanced risk assessment.
\end{tabular}                             & \begin{tabular}[c]{@{}l@{}}  Data freshness and accuracy \\ownership and privacy\\\end{tabular}  &
\\ \cline{1-3}
Space-based cloud computing                                                                            & \begin{tabular}[c]{@{}l@{}} IoT data analytics, multi-user DT, data archiving,\\ AI model training, redundancy and backup in space,\\ immunity to natural disasters occurring on Earth \end{tabular}   & \begin{tabular}[c]{@{}l@{}}Resource-limited devices,\\ security and reliability\end{tabular}      &     \\ \cline{1-3}
Neuromorphic computing   & \begin{tabular}[c]{@{}l@{}}Energy-efficient computation, autonomous device learning,\\ dynamic data analysis and decision-making,\\  self-adaptive networks, AI model integration\end{tabular}      & \begin{tabular}[c]{@{}l@{}}Computational complexity,\\ resource-limited devices\end{tabular} &    \\ \cline{1-3}
Quantum cryptography                                                                       & \begin{tabular}[c]{@{}l@{}}Unconditionally secure communication, quantum-secure DTs,\\quantum key distribution, post-quantum security,\\ protection against eavesdropping\end{tabular}   & Security and reliability &    \\ \cline{1-3} {Quantum computing}    & \begin{tabular}[c]{@{}l@{}}Solving complex optimization problems, parallel processing\\ quantum machine learning, AI model acceleration\\ 
complex simulation and modeling in NTNs \end{tabular}  & Computational complexity    &        \\ \cline{1-3}
Quantum sensing    &  \begin{tabular}[c]{@{}l@{}} Highly accurate measurements, enhanced IoT sensing,\\environmental monitoring, ultra-precise positioning,\\ remote sensing, precision time synchronization
\end{tabular}  & Data freshness and accuracy     &    \\ \cline{1-3}
\begin{tabular}[c]{@{}l@{}}Open radio access network (O-RAN)\end{tabular}               & \begin{tabular}[c]{@{}l@{}}Flexible deployment, multi-vendor interoperability,\\ real-time configurations, service-specific network slices \\ open interfaces, virtualized network function (VNF) integration\end{tabular}                                    & \begin{tabular}[c]{@{}l@{}}Interoperability and \\ standardization\end{tabular}  &   \\ \cline{1-3}
\end{tabular}
\end{table*}

\subsection{Narrow-Band Internet of things (NB-IoT)}
IoT serves as a cornerstone technology for NTN-DTs by enabling real-time data collection, remote monitoring, and predictive maintenance, all of which enhance the precision and functionality of digital twin models. Through IoT, network operators gain critical insights, enabling more effective and efficient network management. Recognizing its importance, the \ac{3GPP} organization in Release-17 has outlined guidelines for implementing narrowband IoT (NB-IoT) over satellites \cite{Sciddurlo2022}. NB-IoT, known for its low power consumption, is particularly suited for battery-operated devices, facilitating data collection from remote and resource-constrained assets in NTNs. Its adoption is instrumental in realizing DT-NTNs, offering network operators valuable tools to improve network management and DT-NTN accuracy.

In this context, while real-time data transmissions from NB-IoTs is crucial for the success of DT-NTNs, achieving it can be challenging, particularly in scenarios where delays in data collection and transmission are unavoidable. The impact of these delays depends on their magnitude and the specific DT-NTN function they affect. Despite such challenges, DT-NTNs can still deliver significant value by leveraging predictive capabilities and robust algorithms, ensuring operational effectiveness even when delays occur. The benefits vary with the extent of delays:
\begin{itemize}
	\item Minimal Delays (milliseconds): These minor delays typically have negligible impact, allowing the DT-NTN to provide near real-time insights for most applications.
	
	\item Moderate Delays (seconds): For functions such as network optimization or anomaly detection, a delay of a few seconds is often acceptable. The DT-NTN can still maintain an accurate and actionable overview.

	\item Significant Delays (minutes or more): Substantial delays can hinder the DT-NTN's effectiveness for time-sensitive applications. For example, real-time UAV course corrections might be compromised, reducing the twin's ability to respond promptly.
\end{itemize}

Several strategies can mitigate the impacts of delays in DT-NTNs. For instance, leveraging historical data and machine learning can help predict network behavior, compensating for delays in real-time data. This enables the system to maintain situational awareness and take proactive measures. Additionally, instead of transmitting every data point, the system can periodically send summarized insights, reducing the impact of delays while retaining valuable information. Prioritizing critical data for near real-time transmission ensures that essential functions remain up-to-date despite network constraints.

Moreover, NB-IoT devices can be used for security monitoring, ensuring the safety and integrity of the NTNs. Particularly, DTs can incorporate security measures based on NB-IoT data to respond to potential threats and vulnerabilities. More importantly, IoT technology can act as a bridge between diverse communication standards within NTNs by leveraging specific communication protocols to create a unified platform capable of understanding and translating data from multiple sources.  This synergy enhances the overall adaptability and performance of the NTN ecosystem.

\subsection{Machine Learning Techniques}
DT-NTN models rely on extensive data, which has to thoroughly analyzed to extract insights about physical assets. In this regard, machine learning algorithms are essential for analyzing data collected from IoT and sensor devices deployed across NTNs. These advanced algorithms empower enhanced decision-making, predictive maintenance, and anomaly detection, thereby improving the DT-NTN role in network management and resource allocation. Moreover, machine learning can expedite DT-NTN algorithms, potentially enabling real-time operations and large-scale technology testing. For instance, deep learning techniques have shown a remarkable ability in modeling complex satellite communication networks. Researchers have effectively employed diverse learning approaches to investigate and improve various aspects such as network modeling \cite{Hu2020}, resource optimization \cite{Jiang2020}, and network slicing \cite{Rodrigues2022}. Given these advancements, it is clear that machine learning techniques will play a crucial role in shaping the DT-NTN paradigm. However, this progress will necessitate significantly high computational power with massive datasets to support the complexity and scale of these models \cite{Abdullah2023}.

Moreover, the \textit{modular architecture} of DT technology allows the creation of DTs for each asset component, which can then be seamlessly interconnected into a comprehensive integrated DT. This modularity facilitates process replication, knowledge transfer, and empowers intelligence at the edge, federated learning, and transfer learning that ultimately enhance system resilience. This approach not only helps in avoiding costly redundancies within NTNs but also enables the prediction of potential disruptions by identifying weak points, thus maximizing NTN resilience.

\subsection{Generative AI}
Generative AI models have an intrinsic role in enabling the deployment and functionality of DT-NTNs by addressing the challenges of limited and incomplete data in complex NTN environments. These models are adept at synthesizing realistic and contextually accurate data to fill gaps in datasets \cite{Tao2024}, which is critical given the vast and dynamic nature of NTNs. For instance, generative models can simulate complex scenarios such as atmospheric disturbances, satellite trajectories, and signal interference patterns, creating a comprehensive dataset to enhance the accuracy and adaptability of DT-NTNs. By supplementing real-time data from NB-IoT devices and NTN sensors, generative AI empowers twin systems to predict and respond to network fluctuations more effectively, strengthening their resilience across diverse operational conditions.

Beyond data generation, generative AI also enables advanced data augmentation techniques that directly enhance the performance of DT-NTNs. For example, models like Generative Adversarial Networks (GANs) create synthetic samples that reflect real-world diversity, such as user behavior patterns or service-specific traffic dynamics \cite{Zhang2024}. This augmented data helps train DT-NTNs for scenarios that are rare or difficult to simulate manually, improving their capacity to test and optimize network configurations. Moreover, generative AI facilitates the assessment of network risks, such as latency bottlenecks or resource allocation failures, and enables proactive solutions in NTNs. By leveraging the ability to generate diverse, high-quality data, generative AI ensures that DT-NTNs remain effective tools for adaptive, data-driven management of complex NTN ecosystems.

\subsection{Space-based Cloud Computing}
Cloud platforms are essential for the creation and operation of DT-NTNs due to their ability to provide the necessary computational power and storage capacity to develop, maintain, and run twin models effectively.  Cloud-based architectures provide the flexibility to make rapid adjustments to network resources, ensuring they align with dynamic demands and fluctuations in network traffic. This adaptability is critical for optimizing network performance in the variable and complex environments characteristic of NTNs.

In the context of NTNs, cloud platforms can extend beyond terrestrial infrastructures to include \textit{space-based cloud servers}, which are data centers located in space rather than on Earth. Larger satellites or satellite constellations can function as these data centers, offering storage and processing power to smaller satellites with limited resources \cite{Huang2018}. This setup allows the vast amounts of data generated by IoT and sensor devices in space to be processed and analyzed directly within space-based clouds, enabling real-time insights into the behavior of non-terrestrial systems. The integration of DT-NTNs with space-based clouds facilitates online optimization and reduces the computational burden on small NTN assets, which often have constrained onboard processing capabilities. These assets can efficiently interact with AI models and engage in data-driven learning processes. Furthermore, space-based clouds support scalable and flexible resource management by dynamically allocating computational resources to meet varying network demands, whether for high-throughput applications or latency-sensitive communications. This scalability ensures that NTNs can adapt to the diverse requirements of modern communication services.

\subsection{Neuromorphic Computing}
Neuromorphic computers deviate from the traditional  \textit{von Neumann} architecture commonly found in conventional computers. Instead, they emulate the structure of biological brains with artificial neurons and synapses. This unique structure enables them to execute complex tasks such as pattern recognition and learning in ways similar to the human brain. Neuromorphic computing can be particularly valuable in scenarios involving temporal signals and requiring continual learning from diverse data sources in contexts like NTNs. Specifically, neuromorphic hardware and algorithms excel in handling the complex, dynamic, and adaptive processes, which makes them well-suited for real-time data analysis and decision-making, thereby enhancing DT capabilities for simulating and responding to changing network conditions. 

In satellite communications, several research activities, notably by the European Space Agency (ESA), investigate the application of neuromorphic processors for AI-based applications that demand extensive parallelism and matrix-based operations to enhance satellite payload performance, reliability, and power efficiency.
Neuromorphic technology offers several advantages, notably in \textit{energy efficiency} and \textit{on-device adaptation}-critical attributes for resource-constrained devices in NTNs \cite{bersuker2018}. This technology allows for complex computations with lower power consumption compared to traditional architectures. Moreover, its on-device learning and adaptation feature enables autonomous learning within the NTN, enhancing the self-optimizing capabilities of the network and contributing to the advancement of DTs.

\subsection{Quantum Technologies}

From an implementation perspective, NTN entities are mostly interconnected via free-space optical (FSO) links, which are favored in quantum communications protocols due to negligible background thermal radiation at optical frequencies \cite{Hayder2024}. Interestingly, quantum technologies can address key challenges for DTs, leveraging quantum phenomena like superposition and entanglement to offer unconditionally secure communications, ultra-precise sensing, and promising computing capabilities \cite{trichili2020}.

\textit{Quantum cryptography}, especially \ac{QKD} protocols, offers a practical approach of exchanging keys between different communicating parties with unconditional security. The security of these keys is guaranteed by the fundamental laws of physics and do not rely on any computational hardness assumption. These secure protocols can be employed for both the DT servers and twin signaling, ensuring robust security across the system.

\textit{Quantum sensing} provides precise measurements for parameters crucial in DT representation, including position, velocity, and environmental conditions. Particularly, quantum sensors based on atomic and molecular systems offer ultra-high precision in time and frequency,  enabling detailed analysis of signal propagation, interference dynamics, and noise in NTNs.

\textit{Quantum computing} represents a paradigm shift in computing capabilities and functionalities beyond classical computers. Specifically, classical computers often struggle with large-scale combinatorial optimization problems encountered in NTNs. Quantum computing algorithms offer unprecedented scalability due to the exponential size of the quantum computational space, paving the way for efficient modeling and optimization in the dynamic and complex environments of NTNs \cite{MIMOLEO2023}.



\subsection{Open Radio Access Network (O-RAN)}
O-RAN is a novel architecture that decouples hardware and software components, allowing for flexible deployment of network infrastructure. This decoupling enables interoperability between off-the-shelf hardware from various vendors and openness in software and interfaces. The open interfaces within the O-RAN architecture facilitate multi-vendor interoperability and coexistence across functions, ensuring seamless integration of various components of the DTs and efficient data exchange through standardized interfaces. Thus, data-driven network management solutions can be effectively incorporated into O-RAN owing to the separation of hardware and software components \cite{10024837}.

O-RAN presents significant advantages for DT technology in the context of NTNs. It offers flexible, programmable, and cost-effective solutions for the communication infrastructure of NTNs, providing efficient resource management, reduced network latency, and improved overall DT performance. The openness of software and interfaces in O-RAN simplifies data collection and management, making the modeling of interactions between different network elements less complex. O-RAN's functional partitioning flexibility contributes to a more accurate representation of the physical network in the DT, enhancing its predictive and optimization capabilities. Additionally, DT technology facilitates effective O-RAN deployment by enabling the creation of virtual models for testing and optimizing scenarios before implementation in the physical network.

\section{Case Study: Network Slicing in O-RAN NTNs}
This section presents a case study to showcase the potential application of DT for resource optimization in quality-of-service (QoS)-aware NTN scenarios. Specifically, we investigate the implementation of learning techniques for dynamic resource allocation in O-RAN-based NTNs, aiming to facilitate the coexistence of enhanced mobile broadband (eMBB) and ultra-reliable low-latency communications (URLLC) services\footnote{URLLC addresses scenarios involving intermittent data transmissions with stringent latency and reliability requirements. In the context of this paper, the demanded services may originate from space users in lower orbits, such as CubeSats and nanosats \cite{SIN_2021}, and thus, this represents latency-sensitive traffic services.}. The challenge lies in their divergent requirements: eMBB demands high data rates, while URLLC requires ultra-low latency, making simultaneous optimization challenging \cite{Dazhi2023}. Addressing these conflicting demands requires advanced resource management and network design.

\begin{figure}[!t]
	\centering
	\def\svgwidth{240pt}
	\fontsize{8}{4}\selectfont{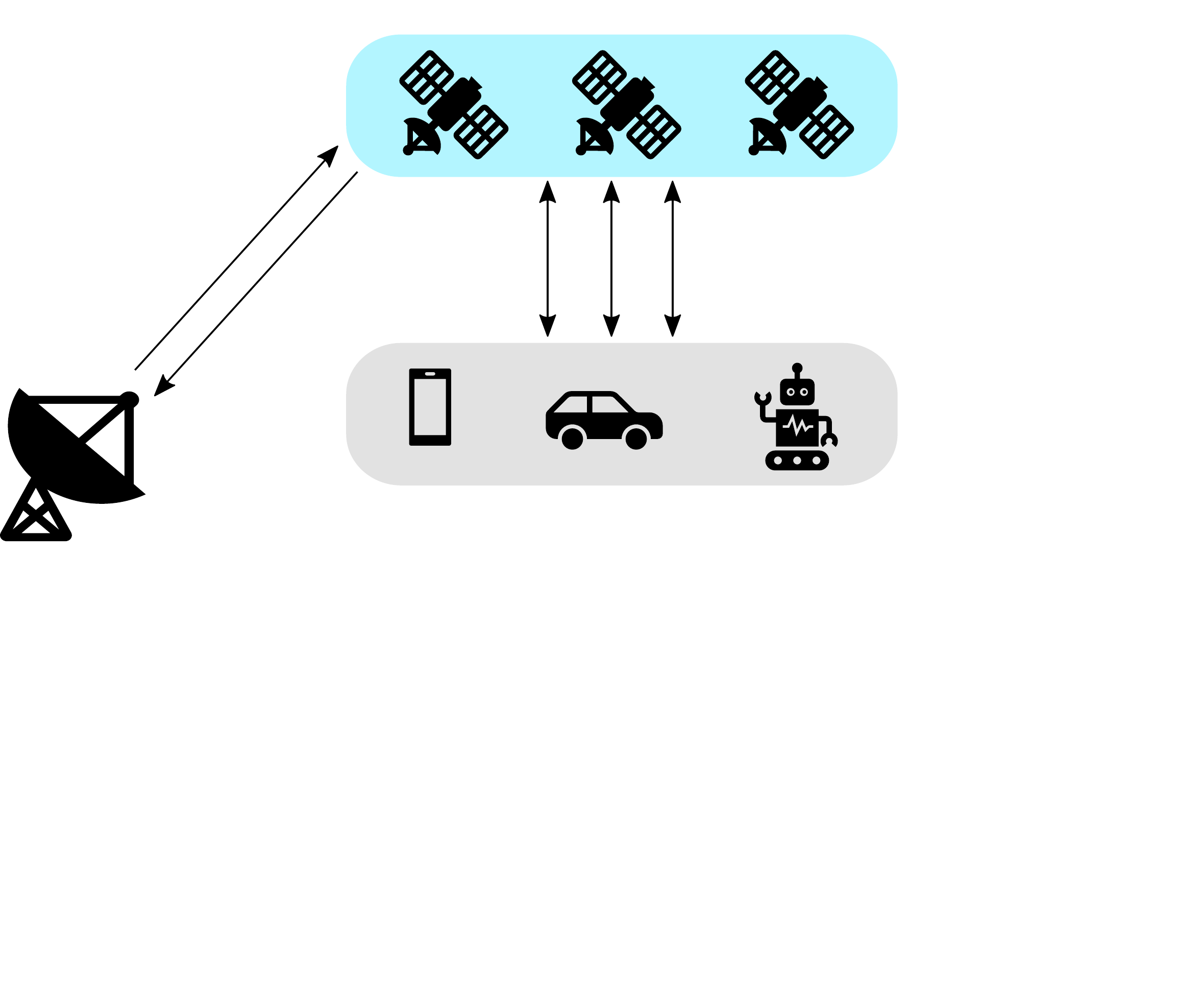}
	\caption{Twin AI-based model for resource allocation in NTNs.}\label{system_model}
\end{figure}

\subsection{Network Scenario}
As depicted in Fig. \ref{system_model}, we consider a constellation of multiple LEO satellites serving multiple distributed users with distinct requirements. In this system, LEO satellites collect network information, including channel states, network traffic, and QoS requirements, and then send the collected data to the non-Real Time Radio access network Intelligent Controller (non-RT RIC) located at a cloud server through a gateway. A DT is constructed based on the collected network information, which will be updated over time to keep synchronizing with the physical network. A learning model based on deep neural networks (DNNs) is installed at the non-RT RIC and trained by interacting with the DT to improve the spectral efficiency while satisfying the QoS requirements of each service. Specifically, the QoS for eMBB services is characterized by a defined minimum data rate threshold, whereas the QoS for URLLC is determined based on the outage probability.

The resource allocation decisions are then performed based on the trained models and sent to both the DT and physical network. The training unit keeps providing decisions and collecting data from the DT to further improve training accuracy and provide better responses to the network dynamic. In the DNN architecture, we use a model consisting of three hidden layers: the first hidden layer is configured with $600$ neurons, the second with $300$ neurons, and the third contains $250$ neurons. The number of neurons in the input layer is aligned with the considered network information, including channel states, network traffic, and QoS requirements. Furthermore, the number of neurons in the output layer corresponds to the size of the resource allocation matrix. We utilize the Rectified Linear Unit (ReLU) activation function for the hidden layers, while the output layer employs the Softmax function.

 \label{key}

\subsection{Results Discussion}
\begin{figure}
    \centering
    \includegraphics[width=0.8\linewidth]{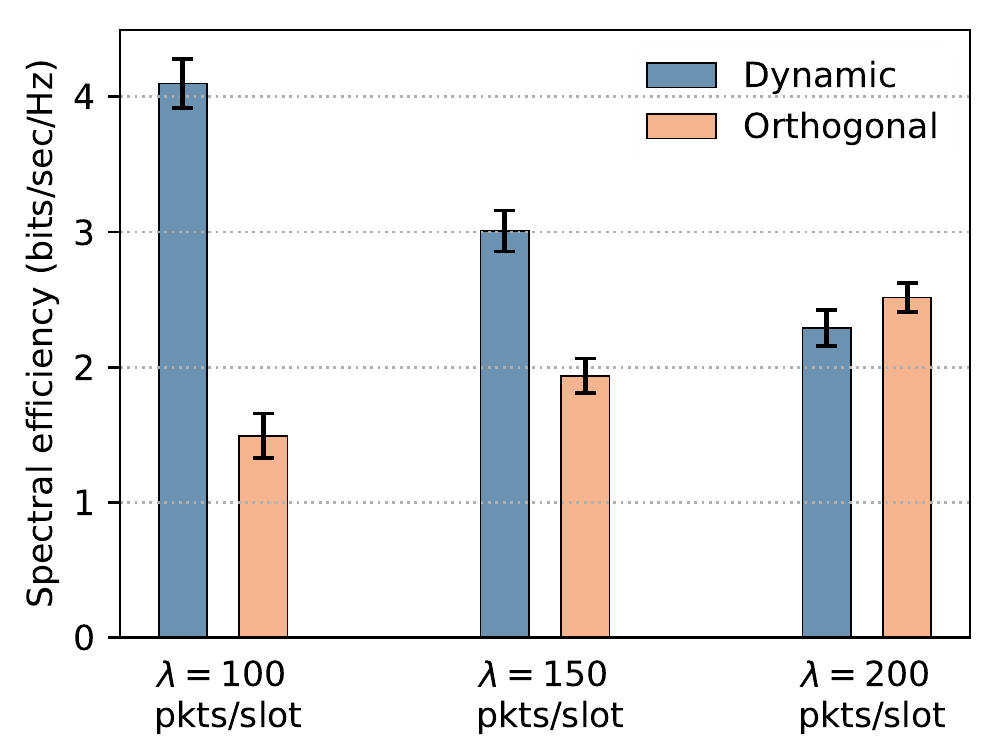}
	\caption{\small Downlink spectral efficiency.}
	\label{spect_eff}
\end{figure}

Fig. \ref{spect_eff} shows the downlink spectral efficiency for different settings of URLLC traffic rate ($\lambda$). Here, spectral efficiency is obtained as the sum data rate of eMBB and URLLC users divided by the system bandwidth. A fully buffered traffic model is considered for eMBB users. We compare the dynamic resource allocation-based approach to the static orthogonal method, where pre-determined fixed resources are assigned to each service. The findings reveal that the AI-based dynamic resource allocation approach provides better resource utilization compared to the orthogonal technique. Nevertheless, in the case of heavy URLLC traffic, the orthogonal method may perform slightly better than the dynamic approach since most of the resources assigned to URLLC users will be utilized. As indicated in Fig. \ref{spect_eff}, the dynamic approach yields approximately $60\%$ higher spectral efficiency compared to the orthogonal method when $\lambda=100$ packet/time slot. This gap decreases with increasing URLLC traffic rate as more eMBB resources will be allocated to serve the higher priority URLLC traffic, impacting the overall eMBB data rate. Lastly, it is noteworthy that increasing the URLLC traffic rate to $\lambda=200$ packets/ time slot elevates the spectral efficiency of the orthogonal approach to about 2.2 bits/second/Hz and reduces the spectral efficiency of the dynamic approach to roughly 2.1 bits/second/Hz.

\begin{figure}
    \centering
    \includegraphics[width=0.9\linewidth, height=2.3in]{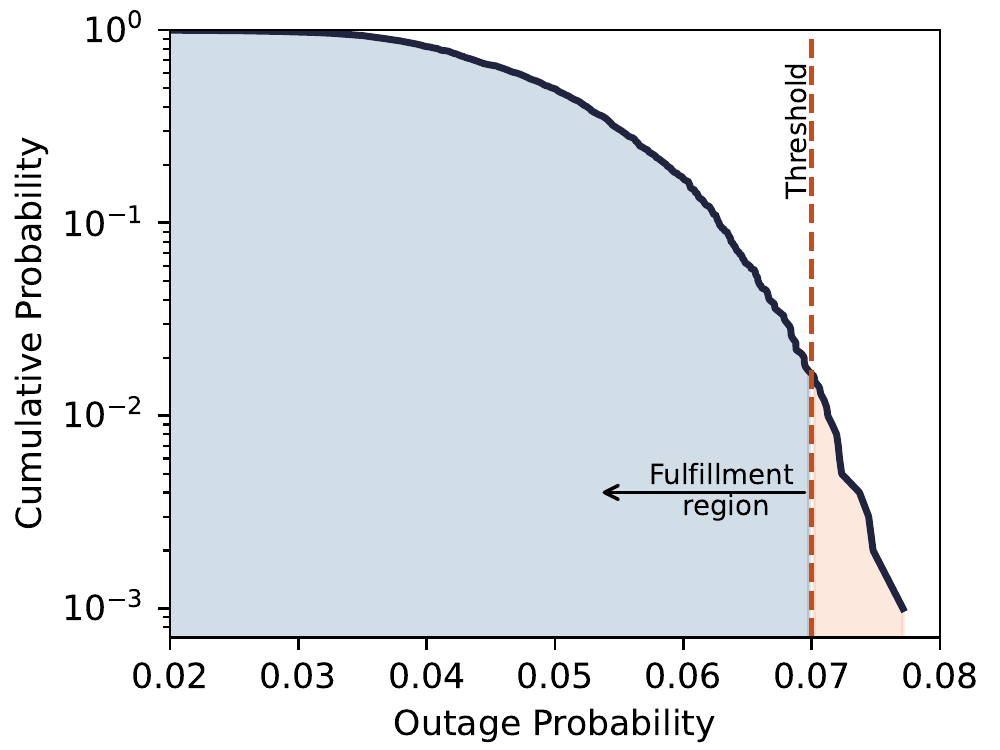}
	\caption{\small URLLC outage probability.}
	\label{outage_probability}
\end{figure}

Fig. \ref{outage_probability} depicts the Cumulative Distribution Function (CDF) of URLLC reliability defined in terms of the outage probability $\textsf{Pr}\left[R_u(t)\leq \zeta \lambda(t)\right]\leq\epsilon_{\mathsf{max}}$, where $R_u(t)$ is the obtained sum data rate of URLLC users at time slot $t$, $\zeta$ represents the URLLC packet size and $\epsilon_{\mathsf{max}}$ denotes the maximum threshold of the outage probability. The results obtained at $\epsilon_{\mathsf{max}}=0.07$ and $\zeta=32$ bytes, while the value of $\lambda$ varies over time slots. Notably, the cumulative probability that the outage probability exceeds the threshold $\epsilon_{\mathsf{max}}$ is around $0.02$. This outcome is due to the fact that the dynamic scheduling algorithm prioritizes critical URLLC traffic by allocating resources from eMBB users over time slots, considering the stochastic network dynamics.

In general, these findings underscore the potential  advantages of incorporating AI models within DTs for enhancing the performance of NTNs, ensuring the required reliability, and meeting the diverse QoS requirements.

\section{Conclusions}\label{sec:conclusions}
This article advocates for the transformative role of DT technology in enhancing the control and management of NTNs. We have presented the vision of integrating DTs into NTNs, and discussed the key deployment challenges including the need for up-to-date and accurate data, high computational power, reliable interconnections, interoperability issues, and secure data procedures.
Various enabling technologies have been explored  to facilitate the integration and address these challenges. In addition, we have presented a case study  that leverages a learning model for network slicing in an O-RAN NTN architecture.
In short, this article has introduced  various research and development aspects regarding the integration of DT technology in NTNs. This may serve as a catalyst for further exploration, potentially revolutionizing network control and management in non-terrestrial communication systems.

\section{Acknowledgments}
This research was funded in whole by the Luxembourg National Research Fund (FNR) under the project SmartSpace, grant reference [C21/IS/16193290]. 

\bibliographystyle{IEEEtran}
\bibliography{IEEEabrv,References}

\renewenvironment{IEEEbiography}[1]
{\IEEEbiographynophoto{#1}}
{\endIEEEbiographynophoto}

\end{document}